\documentclass[showpacs,aps,graphicx]{revtex4}
\usepackage{graphicx}
\begin{document}
\title{Atomic entanglement purification using photonic Faraday rotation} 

\author{Yu-Bo Sheng,$^{1,2}$\footnote{Email address:
shengyb@njupt.edu.cn}  Sheng-Yang Zhao$^{1,2}$, Jiong Liu$^{1,2}$}

\address{ $^1$Institute of Signal Processing  Transmission, Nanjing
University of Posts and Telecommunications, Nanjing, 210003,  China\\
$^1$Key Lab of Broadband Wireless Communication and Sensor Network
 Technology,Nanjing University of Posts and Telecommunications, Ministry of
 Education, Nanjing, 210003,
 China\\}

\begin{abstract}
We describe an entanglement purification protocol (EPP) for atomic entangled pair using photonic Faraday rotation.  It is shown that after the two single photons
input-output process in cavity QED,  the high quality entangled atomic state can be obtained from the
low quality mixed entangled atomic states.  Different from
other EPPs, the two pairs of mixed states do not need to intact directly. As the photonic Faraday rotation works on the low-Q cavities, this EPP is useful in both quantum communication and computation.
\end{abstract}
\maketitle

\section{Introduction}
Entanglement is the important resource in current quantum information processing \cite{rmp}.
For example, quantum teleportaton \cite{teleportation,teleportation1}, quantum dense coding \cite{densecoding}, quantum key distribution \cite{Ekert91}, quantum state
sharing \cite{QSS1,QSS2,QSS3}, and quantum direct secure communication \cite{QSDC1,QSDC2,QSDC3}, they all require the maximally entangled state to
set up the quantum channel. However, the entanglement will inevitably interact with the
environment and it will degrade during the process of distribution and storage in the solid system.

Entanglement purification is a powerful way for distilling the high quality mixed entangled states in a small ensemble from the low quality
mixed states in a large ensembles. In the early works of entanglement purification, they usually used the CNOT gates or
similar logic gates, such as the entanglement purification protocol (EPP) proposed by Bennett \emph{et. al.} \cite{Bennett}, the improved
 EPP proposed by Deustch \emph{et. al.} \cite{Deutsch}, the EPP for multipartite entangled systems \cite{murao}, and so on \cite{highdemension}.  In long-distance quantum communication,
the photons as the flying qubits are the good candidate for carrying quantum information. So there are many EPPs based on the
photons.  For example, In 2001, Pan \emph{et al.} proposed the EPP with the linear optics \cite{pan1}. In 2003, they realized their protocol in experiment \cite{pan2}.
In 2002, Simon and Pan proposed an EPP using spatial entanglement \cite{simon}. In 2008, EPP based on the cross-Kerr nonlinearity was proposed \cite{shengpra1}.
There are also other EPPs for photonic systems \cite{shengpra2,shengpra3,shengepjd,dengpra,lipra}. Actually, the flying qubits distribute the  entangled states between distance locations, and they should be stored into the solid  memory systems
for the further application.  So there are other EPPs for such solid systems \cite{caozl,fengpra,purificationatom,wangpra,shengpla,dengqip,wangqip}. In 2005, Cao\emph{ et al.} proposed an EPP for  arbitrary unknown ionic states via linear optics \cite{caozl}. In the same year, Feng \emph{et. al.} proposed an EPP with charge detection \cite{fengpra}.  In 2006, Reichle \emph{et. al.} reported their experimental report for purification of two-atom entanglement \cite{purificationatom}. In 2011, Wang \emph{et al.} proposed an EPP for hybrid entangled state using
quantum-dot and microcavity coupled system \cite{wangpra}.

On the other hand,  over the pase decades, cavity quantum electrodynamics (QED) has become an important platform for understanding the basic principle
of quantum mechanics and quantum information processing \cite{rmpqed}. Some excellent experiments, such as the coupling of the trapped atoms with cavity \cite{trapped}, generation
of single photons \cite{singlephoton}, conditional logic gate \cite{logicgate}  have been reported. However, the quantum information processing based on the high-Q cavities
and strong coupling to the confined atoms are still the challenge in current techniques. Interestingly, in 2009, An \emph{et al.} discussed the
quantum information processing with a single photon by an input-output process with low-Q cavities \cite{fengmang1}. It is shown that the differently polarized
photons can obtain different phase shift after they interacting with an atom trapped in a low-Q cavity. It is so called the Faraday rotation. Based on the
Faraday rotation, the protocols for entanglement generation \cite{entanglementgeneration}, quantum logic gates \cite{logicgate1}, quantum teleportation \cite{fengmang3}, controlled teleportation \cite{cteleportationqip},
entanglement swapping \cite{swapping} and entanglement concentration for atoms were proposed \cite{atomconcentration}.

Inspired by above works with Faraday rotation \cite{fengmang1,entanglementgeneration,logicgate1,fengmang3,cteleportationqip,swapping,atomconcentration}, we discuss the EPP for entangled atomic systems. The information of the entangled states are encoded
in lambda configured three-level atoms trapped inside coupled cavities by optical fibers. Through the two single photons input-output
process in cavity QED, we can extract the high fidelity mixed entangled atomic state from two pairs of low fidelity mixed entangled states.
The distinct feature of this EPP is that the two pairs of atomic entangled states do not intact directly and can be purified by photonic Faraday rotation.
Furthermore, this EPP works on the low-Q cavities  which lead this EPP
feasible in current experimental condition.

This paper is organized as follows: In Sec. 2, we briefly discuss the basic principle of the photonic Faraday rotation.
In Sec. 3, we explain our EPP. In Sec. 4, we make a discussion and summary.

\section{Basic theoretical model of phonic Faraday rotation}
In this section, we will explain the basic theoretical
model for our protocol. As shown in Fig. 1, the cavity coherently
interacts with a trapped two-level atom. The atom has two degenerate round states $|g_{L}\rangle$
and $|g_{R}\rangle$.  The transitions between two ground states and the excited state $|e\rangle$ are assisted with
left-circularly (L) and right-circularly (R) polarized photon.
 The Hamiltonian of this system is written as \cite{cteleportationqip,swapping}
\begin{eqnarray}
H=\sum_{j=L,R}[\frac{\hbar\omega_{0}\sigma_{jz}}{2}+\hbar\omega_{c}a^{\dagger}_{j}a_{j}]
+\hbar\lambda\sum_{j=L,R}(a_{j}^{\dagger}\sigma_{j-}+a_{j}\sigma_{j+})+H_{R},
\end{eqnarray}
with
\begin{eqnarray}
H_{R}&=&H_{R0}+i\hbar[\int_{-\infty}^{\infty}d\omega\sum_{j=L,R}\alpha(\omega)(b_{j}^{\dagger}(\omega)a_{j}+b_{j}(\omega)a^{\dagger}_{j})]\nonumber\\
&+&\int_{-\infty}^{\infty}d\omega\sum_{j=L,R}\overline{\alpha}(\omega)(c_{j}^{\dagger}\sigma_{j-}+c_{j}\sigma_{j+}).
\end{eqnarray}
Here $a^{\dag}$ and $a$ are the creation and annihilation operators of the cavity field and the frequency of the cavity
field is $\omega_{c}$. The $\omega_{0}$ is the atomic frequency. The $\sigma_{R-}$, $\sigma_{R+}$, $\sigma_{L-}$  and $\sigma_{R+}$
are the lowering and raising operators of the transition R(L), respectively. $H_{R0}$ means the Hamiltonian of the free reservoirs. The field and the atomic reservoirs
are given by
\begin{eqnarray}
H_{RC}=\hbar\int^{\infty}_{-\infty}d\omega b_{j}^{\dagger}b_{j},\nonumber\\
H_{RA}=\hbar\int^{\infty}_{-\infty}d\omega c_{j}^{\dagger}c_{j}.
\end{eqnarray}
 The coupling amplitudes $\alpha=\sqrt{\kappa/2\pi}$ and  $\overline{\alpha}=\sqrt{\gamma/2\pi}$, respectively. $\kappa$ and $\gamma$ are the cavity-field
 and atomic damping rates. The $b^{\dagger}_{j}$ and $c^{\dagger}_{j}$ ($b_{j}$ and $c_{j}$)are the creation (annihilation) operators of the reservoirs.
\begin{figure}[!h]
\begin{center}
\includegraphics[width=8cm,angle=0]{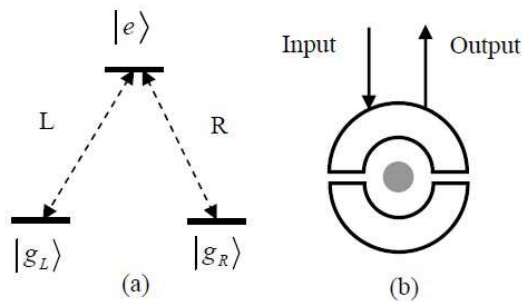}
\caption{(a) The relevant three-level atomic structure confined in a cavity. $|g_{R}\rangle$ and $|g_{L}\rangle$ are the two degenerated ground states
which couple with a right ($|R\rangle$) and a left ($|L\rangle$) polarization photon, respectively. (b) Atomic configuration of the three-level atom trapped in the low-Q cavity. Interaction between the atom and a single-photon pulse propagating input-output in the low-Q cavity according to the photonic Faraday rotation.}
\end{center}
\end{figure}
 We consider a single photon pulse with frequency $\omega_{P}$ input in the optical cavity as shown in Fig. 1 (b).
In the rotating frame with respect to the frequency of the input pulse, the Heisenberg equation of the cavity mode driven by the corresponding cavity
input state can be written as
\begin{eqnarray}
\dot{a}_{j}(t)&=&-[i(\omega_{c}-\omega_{p})+\frac{\kappa}{2}]a_{j}(t)-\lambda\sigma_{j-}(t)-\sqrt{\lambda}a_{in,j}(t),\nonumber\\
\dot{\sigma}_{j-}(t)&=&-[i(\omega_{c}-\omega_{p})+\frac{\gamma}{2}]]a_{j-}(t)-\lambda\sigma_{iz}(t)a_{j}(t)+\sqrt{\gamma}\sigma_{z}(t)b_{in,j}(t).
\end{eqnarray}
The input and output fields of the cavity are related by the intracavity field as
\begin{eqnarray}
a_{out,j}(t)=a_{in,j}(t)+\sqrt{\kappa}a_{j}(t).
\end{eqnarray}
 We denote $r(\omega)\equiv\frac{a_{out,j(t)}}{a_{in,j(t)}}$ as the reflection coefficient of the atom-cavity system.
Here, assuming that we have a weak excitation by the single-photon pulse on the atom initially prepared in the ground state, which means that
we keep $\langle\sigma_{z}\rangle=-1$ through out operation. Using the adiabatic approximation, we can obtain the input-output relation of the cavity  field
in the form of \cite{entanglementgeneration,cteleportationqip,swapping}
\begin{eqnarray}
r(\omega_{p})=\frac{[i(\omega_{c}-\omega_{p})-\frac{\kappa}{2}][i(\omega_{0}-\omega_{p})+\frac{\gamma}{2}]
+\lambda^{2}}{[i(\omega_{c}-\omega_{p})+\frac{\kappa}{2}][i(\omega_{0}-\omega_{p})+\frac{\gamma}{2}]+\lambda^{2}}.\label{reflect1}
\end{eqnarray}
From Eq. (\ref{reflect1}), if $\lambda=0$, above equation for an empty cavity can be written as
\begin{eqnarray}
r_{0}(\omega_{p})=\frac{i(\omega_{c}-\omega_{p})-\frac{\kappa}{2}}{i(\omega_{c}-\omega_{p})+\frac{\kappa}{2}}.
\end{eqnarray}
So if the atom initially is in the state $|g_{L}\rangle$ and a photon with the $|L\rangle$ (left circular polarization)
entrances into the cavity, the $|g_{L}\rangle\leftrightarrow|e\rangle$ is due to the coupling of the cavity mode $a_{L}$
and the $|L\rangle$ photon. After the photon is reflected to the output mode, the state becomes $|L\rangle\rightarrow r(\omega_{p})|L\rangle\approx e^{i\theta}|L\rangle$. The output state leads a phase shift which is determined by the parameter values of the cavity. Interestingly, if the
atom initially is in the state $|g_{L}\rangle$ and a photon with the the $|R\rangle$ (right circular polarization) entrances into the cavity,
it only senses the empty cavity. So after the photon is reflected and in the output mode, it becomes
$|R\rangle\rightarrow r_{0}(\omega_{p})|R\rangle\approx e^{i \theta_{0}}|R\rangle$. Therefore, if the initial photon is in the linearly polarized
state $\alpha|L\rangle+\beta|R\rangle$ and the atom is in the state  $|g_{L}\rangle$,  after the photon passing through the cavity and being reflected, it  becomes
\begin{eqnarray}
\alpha|L\rangle+\beta|R\rangle\rightarrow \alpha e^{i\theta}|L\rangle+\beta e^{i \theta_{0}}|R\rangle.
\end{eqnarray}
On the other hand, if the atom is in the state  $|g_{R}\rangle$, we can also obtain the output mode of the photon as
\begin{eqnarray}
\alpha|L\rangle+\beta|R\rangle\rightarrow \alpha e^{i \theta_{0}}|L\rangle+\beta e^{i\theta}|R\rangle.
\end{eqnarray}
The $\Delta\Theta_{F}=\theta-\theta_{0}$ or $\Delta\Theta_{F}=\theta_{0}-\theta$ is called the Faraday rotation.

If we consider the cases that $\omega_{0}=\omega_{c}$, $\omega_{p}=\omega_{c}-\frac{\kappa}{2}$, and $\lambda=\frac{\kappa}{2}$, we can
obtain $\theta=\pi$ and $\theta_{0}=\frac{\pi}{2}$. The evolution of the photon and atom can be written as
\begin{eqnarray}
|L\rangle|0\rangle\rightarrow-|L\rangle|0\rangle, |R\rangle|0\rangle\rightarrow i|R\rangle|0\rangle,\nonumber\\
|L\rangle|1\rangle\rightarrow i|L\rangle|1\rangle, |R\rangle|1\rangle\rightarrow -|R\rangle|1\rangle.\label{relation}
\end{eqnarray}
Here $|g_{L}\rangle\equiv|0\rangle$ and $|g_{R}\rangle\equiv|1\rangle$.

\section{EPP using Faraday rotation}
Now we start to explain our protocol with a simple example.
From Fig. 2, suppose that the entangled pairs of two atoms $a_{1}$ and $b_{1}$, $a_{2}$ and $a_{2}$
are in the same mixed state written as
\begin{eqnarray}
\rho_{ab}=F|\phi^{+}\rangle_{ab}\langle\phi^{+}|+(1-F)|\psi^{+}\rangle_{ab}\langle\psi^{+}|.\label{mixed}
\end{eqnarray}
Here $|\phi^{+}\rangle_{ab}=\frac{1}{\sqrt{2}}(|0\rangle_{a}|0\rangle_{b}+|1\rangle_{a}|1\rangle_{b})$, and
 $|\psi^{+}\rangle_{ab}=\frac{1}{\sqrt{2}}(|0\rangle_{a}|1\rangle_{b}+|1\rangle_{a}|0\rangle_{b})$.
\begin{figure}[!h]
\begin{center}
\includegraphics[width=12cm,angle=0]{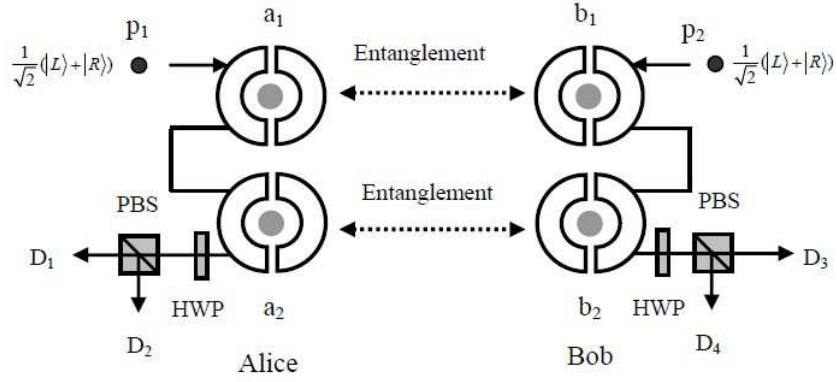}
\caption{A schematic drawing of the
basic element of our EPP. Two pairs of low fidelity of mixed states $a_{1}b_{1}$ and $a_{2}b_{2}$ belong to Alice and Bob.
The two single-photon pulses are sent to the  $a_{1}$ and $b_{1}$, respectively. According the the measurement results of the photons, Alice and Bob
decide to remain or discard the atomic entangled states. }
\end{center}
\end{figure}
From Eq. (\ref{mixed}), the mixed state contains a bit-flip error with the probability of $1-F$.
The two   mixed states $\rho_{a_{1}b_{1}}\otimes\rho_{a_{2}b_{2}}$ can be viewed as the four combinations of pure states: with the probability of $F^{2}$, it is in the state $|\phi^{+}\rangle_{a_{1}b_{1}}\otimes|\phi^{+}\rangle_{a_{2}b_{2}}$, with the equal probability
$F(1-F)$, it is in the state $|\phi^{+}\rangle_{a_{1}b_{1}}\otimes|\psi^{+}\rangle_{a_{2}b_{2}}$ and  $|\psi^{+}\rangle_{a_{1}b_{1}}\otimes|\phi^{+}\rangle_{a_{2}b_{2}}$, with the probability of $(1-F)^{2}$, it is in the state  $|\psi^{+}\rangle_{a_{1}b_{1}}\otimes|\psi^{+}\rangle_{a_{2}b_{2}}$.
Our EPP can be started as follows: first, Alice and Bob both prepare the single photon of the form $\frac{1}{\sqrt{2}}(|L\rangle+|R\rangle)$.
The two single photons are named $p_{1}$ and $p_{2}$, respectively, as shown in Fig. 2.
Then they let their photons entrance into the cavities. After the two photons coupling with the cavities, they will finally
 be output of the cavities and  detected by the two of the four single-photon detectors. If  the detectors D$_{1}$D$_{3}$ or   D$_{2}$D$_{4}$  fire, then they remain
 the atomic mixed states, which is a successful case for purification. Otherwise, if the  detectors D$_{1}$D$_{4}$ or   D$_{2}$D$_{3}$  fire, they will discard the atomic mixed states for it is a failure.
The cross-combinations $|\phi^{+}\rangle_{a_{1}b_{1}}\otimes|\psi^{+}\rangle_{a_{2}b_{2}}$ and  $|\psi^{+}\rangle_{a_{1}b_{1}}\otimes|\phi^{+}\rangle_{a_{2}b_{2}}$ cannot lead the D$_{1}$D$_{3}$ or   D$_{2}$D$_{4}$  fire.
 In detail,  the cross-combination item $|\phi^{+}\rangle_{a_{1}b_{1}}\otimes|\psi^{+}\rangle_{a_{2}b_{2}}$ coupling with the two single photons evolves as
\begin{eqnarray}
&&\frac{1}{\sqrt{2}}(|L\rangle_{p_{1}}+|R\rangle_{p_{1}})\otimes \frac{1}{\sqrt{2}}(|L\rangle_{p_{2}}+|R\rangle_{p_{2}})\otimes|\phi^{+}\rangle_{a_{1}b_{1}}\otimes|\psi^{+}\rangle_{a_{2}b_{2}}\nonumber\\
&=&\frac{1}{4}[(|L\rangle_{p_{1}}+|R\rangle_{p_{1}})\otimes(|L\rangle_{p_{2}}+|R\rangle_{p_{2}})\otimes
(|0\rangle_{a_{1}}|0\rangle_{a_{2}}|0\rangle_{b_{1}}|1\rangle_{b_{2}}+|0\rangle_{a_{1}}|1\rangle_{a_{2}}|0\rangle_{b_{1}}|0\rangle_{b_{2}})\nonumber\\
&+&|1\rangle_{a_{1}}|0\rangle_{a_{2}}|1\rangle_{b_{1}}|1\rangle_{b_{2}}+|1\rangle_{a_{1}}|1\rangle_{a_{2}}|1\rangle_{b_{1}}|0\rangle_{b_{2}}]\nonumber\\
&\rightarrow&[-i(|L\rangle_{p_{1}}-|R\rangle_{p_{1}})\otimes(|L\rangle_{p_{2}}+|R\rangle_{p_{2}})
\otimes(|0\rangle_{a_{1}}|0\rangle_{a_{2}}|0\rangle_{b_{1}}|1\rangle_{b_{2}}-|1\rangle_{a_{1}}|1\rangle_{a_{2}}|1\rangle_{b_{1}}|0\rangle_{b_{2}})\nonumber\\
&-&i(|L\rangle_{p_{1}}+|R\rangle_{p_{1}})\otimes(|L\rangle_{p_{2}}-|R\rangle_{p_{2}})(|0\rangle_{a_{1}}|1\rangle_{a_{2}}|0\rangle_{b_{1}}|0\rangle_{b_{2}}
-|1\rangle_{a_{1}}|0\rangle_{a_{2}}|1\rangle_{b_{1}}|1\rangle_{b_{2}}).\label{cross1}\nonumber\\
\end{eqnarray}
The $|\phi^{+}\rangle_{a_{1}b_{1}}\otimes|\phi^{+}\rangle_{a_{2}b_{2}}$ coupling with the two single photons evolve as
 \begin{eqnarray}
&& \frac{1}{\sqrt{2}}(|L\rangle_{p_{1}}+|R\rangle_{p_{1}})\otimes \frac{1}{\sqrt{2}}(|L\rangle_{p_{2}}+|R\rangle_{p_{2}})\otimes|\phi^{+}\rangle_{a_{1}b_{1}}\otimes|\phi^{+}\rangle_{a_{2}b_{2}}\nonumber\\
&=&\frac{1}{4}[(|L\rangle_{p_{1}}+|R\rangle_{p_{1}})\otimes(|L\rangle_{p_{2}}+|R\rangle_{p_{2}})
\otimes(|0\rangle_{a_{1}}|0\rangle_{a_{2}}|0\rangle_{b_{1}}|0\rangle_{b_{2}}+|0\rangle_{a_{1}}|1\rangle_{a_{2}}|0\rangle_{b_{1}}|1\rangle_{b_{2}})\nonumber\\
&+&|0\rangle_{a_{1}}|1\rangle_{a_{2}}|1\rangle_{b_{1}}|0\rangle_{b_{2}}+|1\rangle_{a_{1}}|1\rangle_{a_{2}}|1\rangle_{b_{1}}|1\rangle_{b_{2}}]\nonumber\\
&\rightarrow&\frac{1}{4}[(|L\rangle_{p_{1}}-|R\rangle_{p_{1}})\otimes(|L\rangle_{p_{2}}-|R\rangle_{p_{2}})
\otimes(|0\rangle_{a_{1}}|0\rangle_{a_{2}}|0\rangle_{b_{1}}|0\rangle_{b_{2}}+|1\rangle_{a_{1}}|1\rangle_{a_{2}}|1\rangle_{b_{1}}|1\rangle_{b_{2}})\nonumber\\
&-&(|L\rangle_{p_{1}}+|R\rangle_{p_{1}})\otimes(|L\rangle_{p_{2}}+|R\rangle_{p_{2}})
\otimes(|0\rangle_{a_{1}}|1\rangle_{a_{2}}|0\rangle_{b_{1}}|1\rangle_{b_{2}})+|1\rangle_{a_{1}}|0\rangle_{a_{2}}|1\rangle_{b_{1}}|0\rangle_{b_{2}})].\label{cross2}\nonumber\\
\end{eqnarray}

After the two photons reflect from two cavities respectively, both of each perform a Hadamard  operation on their photons and let them pass through
the polarization beam splitters (PBSs) which transmit the $|L\rangle$ photon and reflect the $|R\rangle$ photon, respectively. The Hadamard operation can be described as
\begin{eqnarray}
|L\rangle\rightarrow\frac{1}{\sqrt{2}}(|L\rangle+|R\rangle),\nonumber\\
|R\rangle\rightarrow\frac{1}{\sqrt{2}}(|L\rangle-|R\rangle).
\end{eqnarray}
Finally, the two photons are detected by single-photon detectors. From  Eq. (\ref{cross1}),
the polarization of the photons are different. They cannot lead the D$_{1}$D$_{3}$ or   D$_{2}$D$_{4}$  fire. On the other hand, If the initial state is $|\psi^{+}\rangle_{a_{1}b_{1}}\otimes|\phi^{+}\rangle_{a_{2}b_{2}}$, they can obtain the similar result analogy with Eq. (\ref{cross1}).   So the cross-combination items can be
eliminated automatically.

 Therefore, by selecting the cases that
both Alice and Bob have the same measurement results $|L\rangle$, they will  obtain
\begin{eqnarray}
|\Phi^{+}\rangle=\frac{1}{\sqrt{2}}(|0\rangle_{a_{1}}|0\rangle_{a_{2}}|0\rangle_{b_{1}}|0\rangle_{b_{2}}+|1\rangle_{a_{1}}|1\rangle_{a_{2}}|1\rangle_{b_{1}}|1\rangle_{b_{2}}),
\end{eqnarray}
with the probability of $F^{2}$, and
\begin{eqnarray}
|\Psi^{+}\rangle=\frac{1}{\sqrt{2}}(|0\rangle_{a_{1}}|0\rangle_{a_{2}}|1\rangle_{b_{1}}|1\rangle_{b_{2}}+|1\rangle_{a_{1}}|1\rangle_{a_{2}}|0\rangle_{b_{1}}|0\rangle_{b_{2}}),
\end{eqnarray}
with the probability of $(1-F)^{2}$.
On the other hand, if both the measurements results are $|R\rangle$, they will  obtain
\begin{eqnarray}
|\Phi^{+}\rangle'=\frac{1}{\sqrt{2}}(|0\rangle_{a_{1}}|1\rangle_{a_{2}}|0\rangle_{b_{1}}|1\rangle_{b_{2}}+|1\rangle_{a_{1}}|0\rangle_{a_{2}}|1\rangle_{b_{1}}|0\rangle_{b_{2}}),
\end{eqnarray}
with the probability of $F^{2}$, and
\begin{eqnarray}
|\Psi^{+}\rangle'=\frac{1}{\sqrt{2}}(|0\rangle_{a_{1}}|1\rangle_{a_{2}}|1\rangle_{b_{1}}|0\rangle_{b_{2}}+|1\rangle_{a_{1}}|0\rangle_{a_{2}}|0\rangle_{b_{1}}|1\rangle_{b_{2}}),
\end{eqnarray}
with the probability of $(1-F)^{2}$. Finally, they perform the Hadamard operation on the atoms $a_{2}$ and $b_{2}$, which will lead
\begin{eqnarray}
|0\rangle\rightarrow\frac{1}{\sqrt{2}}(|0\rangle+|1\rangle),\nonumber\\
|1\rangle\rightarrow\frac{1}{\sqrt{2}}(|0\rangle-|1\rangle).
\end{eqnarray}
So if the measurement results is $|0\rangle|0\rangle$ or $|1\rangle|1\rangle$, they will obtain the new mixed state in $a_{1}b_{1}$ with
\begin{eqnarray}
\rho_{a_{1}b_{1}}=F'|\phi^{+}\rangle_{a_{1}b_{1}}\langle\phi^{+}|+(1-F')|\psi^{+}\rangle_{a_{1}b_{1}}\langle\psi^{+}|.\label{newmixed}
\end{eqnarray}
Here $F'=\frac{F^{2}}{F^{2}+(1-F)^{2}}$. If $F>\frac{1}{2}$, $F'>F$.
In this way, they can purify the bit-flip error.
On the other hand, if the measurement result is $|0\rangle|1\rangle$ or $|1\rangle|0\rangle$, they will obtain
the $|\phi^{-}\rangle_{a_{1}b_{1}}$ with the same fidelity $F'$. Here $|\phi^{-}\rangle_{a_{1}b_{1}}=\frac{1}{\sqrt{2}}(|0\rangle_{a_{1}}|0\rangle_{b_{1}}-|1\rangle_{a_{1}}|1\rangle_{b_{1}})$. In this case, one of the parities say Alice or Bob should
perform a phase-flip operation on her/his atom to obtain the state in Eq. (\ref{newmixed}).
Once the bit-flip error can be purified, the phase-flip error can also be purified, for
the phase-flip error can be converted into the bit-flip error with the Hadamard operation and it can be purified in a next round.
That is, if a phase-flip error is existed, the mixed state can be written as
 \begin{eqnarray}
\rho'_{ab}=F|\phi^{+}\rangle_{ab}\langle\phi^{+}|+(1-F)|\phi^{-}\rangle_{ab}\langle\phi^{-}|.\label{mixed2}
\end{eqnarray}
Before they starting the EPP, they first perform the Hadamard operations on their own atoms. The
$|\phi^{+}\rangle_{ab}$ does not change after the Hadamard operations while the  $|\phi^{-}\rangle_{ab}$
will become $|\psi^{+}\rangle_{ab}$. They can get the same mixed states as Eq. (\ref{mixed}), which can be purified in
a next round. In this way, they can purify an arbitrary mixed state.

\section{Discussion and summary}
So far, we have briefly described our EPP. During the whole process, the photonic Faraday rotation acts as the key role
in the protocol. As we known, in Ref. \cite{pan1}, the polarization entangled state can be purified with the linear optics,
say PBS. Actually, during their protocol, the function of the PBS is to make a parity check. It is to distinguish the
$|H\rangle|H\rangle$ and $|V\rangle|V\rangle$ states from the $|H\rangle|V\rangle$ and $|V\rangle|H\rangle$ states, according to
the photon number of the output mode of the PBS. Her $|H\rangle$ ($|V\rangle$) are the horizonal (vertical) polarization of
the photon. Similarly, in Ref. \cite{shengpra1}, one can make the parity check by the different phase shift of the coherent, after
the photons and the coherent state coupling with the cross-Kerr nonlinearity. In this paper, the photonic Faraday rotation essentially acts as the same
role of the parity check after the photons passing through the two cavities.  According to Eq. (\ref{relation}), if the photon couples with
the two cavities, we can obtain
\begin{eqnarray}
|L\rangle|0\rangle|0\rangle\rightarrow|L\rangle|0\rangle|0\rangle, |R\rangle|0\rangle|0\rangle\rightarrow -|R\rangle|0\rangle|0\rangle,\nonumber\\
|L\rangle|1\rangle|1\rangle\rightarrow -|L\rangle|1\rangle|1\rangle, |R\rangle|1\rangle|1\rangle\rightarrow |R\rangle|1\rangle|1\rangle,\nonumber\\
|L\rangle|0\rangle|1\rangle\rightarrow-i|L\rangle|0\rangle|1\rangle, |R\rangle|0\rangle|1\rangle\rightarrow -i|R\rangle|0\rangle|0\rangle,\nonumber\\
|L\rangle|1\rangle|0\rangle\rightarrow -i|L\rangle|1\rangle|1\rangle, |R\rangle|1\rangle|0\rangle\rightarrow -i|R\rangle|1\rangle|0\rangle.\label{relation1}
\end{eqnarray}
\begin{figure}[!h]
\begin{center}
\includegraphics[width=8cm,angle=0]{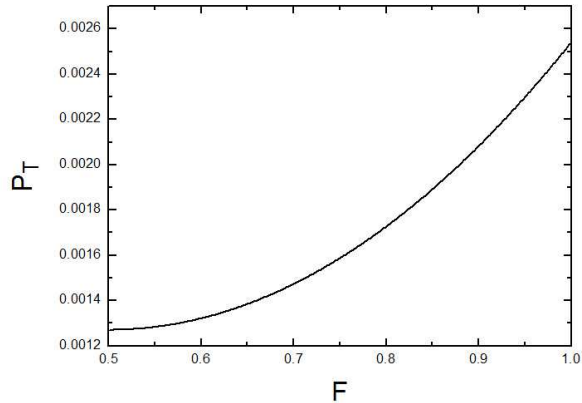}
\caption{The success probability $P_{T}$ is altered with the initial fidelity $F$. We choose $F\in(0.5,1) $ to ensure the initial
mixed state be entangled.}
\end{center}
\end{figure}
From  Eq. (\ref{relation1}), the  $\frac{1}{\sqrt{2}}(|L\rangle+|R\rangle)$ photon will be flipped to  $\frac{1}{\sqrt{2}}(|L\rangle-|R\rangle)$ if the atoms
are in the even parity state  $|0\rangle|0\rangle$ or  $|1\rangle|1\rangle$, while it does not change if the atoms are  in the odd parity state $|0\rangle|1\rangle$ or  $|1\rangle|0\rangle$. In this way, one can make the parity check of the atoms by measuring the photon. Interestingly,
we do not require the atoms to interact directly. So the whole process essentially is the quantum nondemolition (QND) like Refs. \cite{fengpra,shengpla}. In their protocols,
by measuring the charges of the electrons, they can make the parity check, while in this protocol,
with the help of photonic Faraday rotation, we can realize the similar function by measuring the polarization of the photon.

The most advantage for using the photonic Faraday rotation is that the purified high fidelity mixed entangled
state can be remained. It is quite different from the EPP in Ref. \cite{pan1}. In Ref.  \cite{pan1}, after
selecting the four-mode cases, the four photons will be destroyed because of the post-selection principle.
Therefore, they cannot perform the further purification. In this paper, they remain or discard
the atomic entanglement according to the polarization of the photons. So the remained atomic entangled
state can be improved to obtain the higher fidelity in the next purification round.

From above discussion, one can find that  the photon loss may be the main obstacle in realistic experiment.
The photon loss may occur due to cavity mirror absorption and scattering. It can also occurs on
the fiber absorption and the detector inefficiency. From Sec. 3, the successful condition of  purification requires that Alice and Bob pick up the cases that both the photons have the same polarization, according to the classical communication. Therefore, the photon loss does not affect
the fidelity of the mixed state. However, it will decrease the success probability of the whole protocol. Suppose that the efficiency of the single-photon
detector is $\eta_{d}=28\%$ \cite{detectorefficency}, the efficiency of coupling and transmission of each photon through the single-mode fiber is $T_{f}=0.2$, and the transmission of
each photon through the other optical components is $T_{o}=0.9$. We can estimate the total success probability of our EPP as
\begin{eqnarray}
P_{T}=P_{p}\times T^{2}_{f}\times T^{2}_{o}\times\eta^{2}_{d}.
\end{eqnarray}
Here $P_{p}=F^{2}+(1-F)^{2}$. It is the success probability in the ideal case of purification. We calculate the total
success probability altered with the initial fidelity $F$ in Fig. 3. It is shown that if $F=0.8$, $P_{T}\approx 1.7\times 10^{-3}$.
By using the generation rate $1\times10^{-4}s^{-1}$ of the single photons, our EPP can be realized within 0.1 s.
Certainly, in a practical experiment, in order to ensure the two photons reach the two cavities simultaneously, we can use
the spontaneous parametric down-conversion (SPDC) source to substitute the single-photon source. As shown in Ref. \cite{gisin}, a 25-mm-long type-II PPLN waveguide generates both
the input and the auxiliary photon using SPDC source. Then the  pairs are separated by a PBS and can be sent to the different cavities in distant locations.

In conclusion, we have presented an EPP for atomic entanglement with photonic Faraday rotation.
Through the two single photons input-output process in cavity QED, one can obtain one pair of  the high
fidelity of the maximally entangled state from two pairs of low fidelity mixed states. In our EPP, we require
the low-Q cavity, the three-level atom inside the cavity,  some linear optics and the photon detectors  to achieve the task.
The most advantage of this EPP is that the purified atomic entangled state can be remained to repeat to obtain the higher fidelity in
the next purification round. It maybe useful in current long-distance quantum communication and quantum computation.

\section*{ACKNOWLEDGEMENTS}
This work was supported by the National Natural Science Foundation
of China under Grant No. 11104159,   Open Research
Fund Program of the State Key Laboratory of
Low-Dimensional Quantum Physics Scientific, Tsinghua University, Open Research Fund Program of National Laboratory of Solid State Microstructures under Grant No. M25020 and M25022, and the Project
Funded by the Priority Academic Program Development of Jiangsu
Higher Education Institutions.

\end{document}